\documentstyle[twoside,fleqn,espcrc2,epsf]{article}
\renewcommand{\floatsep}{5mm plus 2pt minus 0pt}
\textfloatsep=\floatsep
\intextsep=\floatsep
\title{\bf Gluon excitations of the static quark potential
           and the hybrid quarkonium spectrum\thanks{Talk
           presented by C.~Morningstar and poster
           presented by K.J.~Juge.}}

\author{K.J.~Juge, J.~Kuti, and C.J.~Morningstar\address{Dept.~of Physics,        University of California at San Diego,
        La Jolla, California 92093-0319}}

\begin{document}

\begin{abstract}
A comprehensive determination of the rich, low-lying spectrum
of gluonic excitations in the presence of a static
quark-antiquark pair is presented.  Our results are obtained
from several simulations on anisotropic lattices using an
improved action and a large set of gluonic operators.  The hybrid
quarkonium states are studied in the quenched approximation
using the Born-Oppenheimer expansion and nonrelativistic lattice QCD.
\end{abstract}

\maketitle

\section{Introduction}
Besides conventional hadrons, QCD suggests the existence of states
containing gluonic excitations, such as glueballs and hybrid hadrons.
Although conventional hadrons are reasonably well described by the
constituent quark model, states with excited glue are still poorly
understood.  As experiments begin to focus on the search for glueballs
and hybrid mesons, a better understanding of these states from theory
is needed.  Due to the highly nonperturbative nature of the gluonic
excitations in these states, lattice simulations offer at present the
best means of theoretically probing glueballs and hybrid mesons.

A great advantage in studying hybrid mesons comprised of heavy quarks
is that such systems can be studied not only by direct numerical
simulation, but also using the Born-Oppenheimer (BO) expansion.  In this
approach, the hybrid meson is treated analogous to a diatomic molecule:
the slow heavy quarks correspond to the nuclei and the fast gluon
field corresponds to the electrons\cite{hasenfratz}.
First, one treats the quark $Q$ and
antiquark $\overline{Q}$ as spatially-fixed colour sources and determines the
energy levels of the glue as a function of the $Q\overline{Q}$ separation
$r$; each of these energy levels defines a potential
$V_{Q\overline{Q}}(r)$ adiabatically.  The quark motion is then restored
by solving the Schr\"odinger equation in each of these static potentials.
Conventional quarkonia arise from the lowest-lying static potential;
hybrid quarkonium states emerge from the excited potentials.  Once the
static potentials have been determined (via lattice simulations),
it is a simple matter to determine the complete conventional and hybrid
quarkonium spectrum in the leading Born-Oppenheimer (LBO) approximation.
This is a distinct advantage over meson simulations which yield only
the very lowest-lying states, often with large statistical uncertainties.

Here, we present results for the spectrum of gluonic excitations in the
presence of a static quark-antiquark pair.  Some of these potentials have
been studied before\cite{michael}.  This study is the
first to comprehensively survey the spectrum in SU(3) gauge theory.  Due
to our use of anisotropic lattices and an improved action, we have been
able to determine the static potentials for much larger values of
$Q\overline{Q}$ separation than previously studied.  Using our potentials,
we also determine the hybrid quarkonium spectrum.  Results from a
preliminary nonrelativistic lattice QCD simulation are also presented.

\section{Computation of the potentials}

The first step in the Born-Oppenheimer expansion is the determination
of the rich spectrum of energy levels of the gluons in the presence
of the quark and antiquark, fixed in space some distance $r$ apart.  At
this point in the approximation, the quark and antiquark simply act as
static colour sources.  The gluonic energies (or static potentials)
may be labelled by the magnitude (denoted by $\Lambda$) of the projection
of the total angular momentum of the gluons onto the molecular axis,
by the sign of this projection (chirality or handedness), and by
the behaviour under the combined operations of charge conjugation
and spatial inversion about the midpoint between the quark and the
antiquark.  States with $\Lambda=0,1,2,\dots$
are typically denoted by the capital Greek letters $\Sigma, \Pi,
\Delta, \dots$, respectively.  States which are even (odd) under
the above-mentioned parity--charge-conjugation operation are denoted
by the subscripts $g$ ($u$).  The energy of the gluons is unaffected
by reflections in a plane containing the molecular axis; since such
a reflection interchanges states of opposite handedness, such states
must necessarily be degenerate ($\Lambda$ doubling).  However, this
doubling does not apply to the $\Sigma$ states; $\Sigma$ states which
are even (odd) under a reflection in a plane containing the molecular
axis are denoted by a superscript $+$ $(-)$.  Hence, the low-lying
levels are labelled $\Sigma_g^+$, $\Sigma_g^-$, $\Sigma_u^+$, $\Sigma_u^-$,
$\Pi_g$, $\Pi_u$, $\Delta_g$, $\Delta_u$, and so on.  For convenience,
we use $\Gamma$ to denote these labels in general.

\begin{table}
\setlength{\tabcolsep}{3mm}        
\caption{Simulation parameters, including the coupling $\beta$,
  input aspect ratio $\xi$, approximate lattice spacing $a_s$,
  lattice size, and spatial link smearing parameters ($\lambda$
  and $n_\lambda$ are defined in Ref.~\protect\cite{peardon}).}
\begin{center}
\begin{tabular}{ccccc} \hline
$\beta$ & $\xi$ & $a_s$ (fm) & Lattice & $(\lambda, n_\lambda)$  
    \\ \hline
$2.2$ & 5 & 0.27 & $12^3\times48$ & $(0.10,4)$ \\
      &   &      &                & $(0.20,4)$ \\
      &   &      &                & $(0.30,4)$ \\ \hline
$2.4$ & 5 & 0.23 & $14^3\times56$ & $(0.10,8)$ \\
      &   &      &                & $(0.15,8)$ \\
      &   &      &                & $(0.20,8)$ \\
      &   &      &                & $(0.25,8)$ \\ \hline
$2.6$ & 3 & 0.19 & $10^3\times30$ & $(0.15,8)$ \\
      &   &      &                & $(0.30,8)$ \\ \hline
\end{tabular}
\end{center}
\label{table:simparams}
\end{table}

\epsfverbosetrue
\begin{figure}[t]
\begin{center}
\leavevmode
\epsfxsize=2.9in\epsfbox[0 80 585 576]{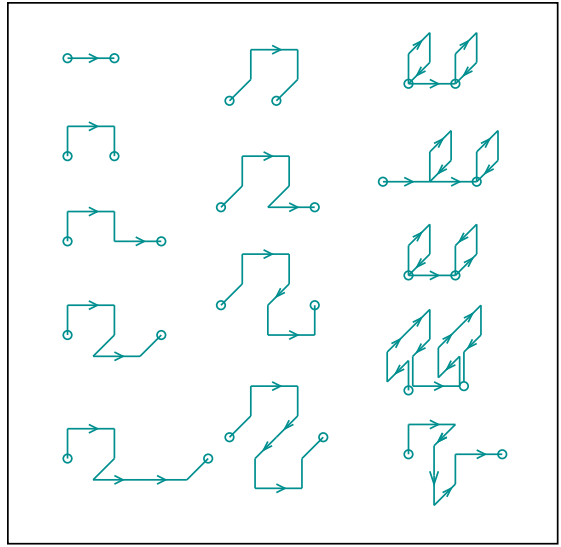}
\end{center}
\caption[figops]{Examples of the paths from the quark to the
  antiquark used to construct the gauge field operators.
\label{fig:ops}}
\end{figure}

Static potentials were extracted from Monte Carlo estimates 
of generalized Wilson loops.  On a starting time slice $t_0$, the
quark and antiquark were fixed at lattice sites a distance $r$
apart.  Several paths along the links of the lattice connecting
the quark and the antiquark were then chosen, and our gluonic operators
$O_i^\Gamma(t_0)$ were defined as linear combinations of the path-ordered
exponentials of the gauge field along these paths. Examples of these paths
are shown in Fig.~\ref{fig:ops}.  The linear combinations were chosen such
that the operators transformed irreducibly under all symmetry operations.
To reduce the mixings of our operators with excited states, the gluonic
operators were constructed from iteratively-smeared spatial links.  We used
the single-link smearing algorithm described in Ref.~\cite{peardon}.
The quark and antiquark then evolved in time, remaining fixed at their
original spatial locations.  To reduce statistical noise, the static
quark propagators, which are simply temporal Wilson lines, were constructed
from thermally-averaged temporal links\cite{thermal}, whenever possible.
The thermal averaging was done using the pseudo-heat-bath method (40 updates).
At some final time slice $t_0\!+\!\tau$, evaluation of the gluonic operators
$O^{\Gamma\dagger}_i(t_0\!+\!\tau)$ then completed the construction of
the Wilson loops $W^\Gamma_{ij}(r,\tau)$.

\epsfverbosetrue
\begin{figure}[t]
\begin{center}
\leavevmode
\epsfxsize=2.9in\epsfbox[18 244 592 690]{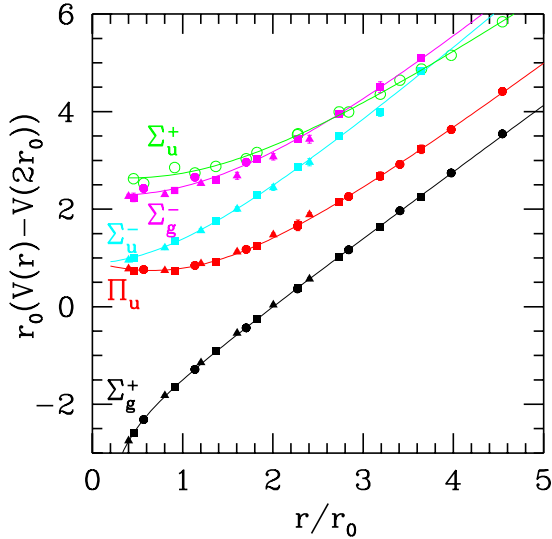}
\end{center}
\caption[figsig]{The static quark potential $V_{\Sigma_g^+}(r)$ and
 some of its gluonic excitations in terms of the hadronic scale
 parameter $r_0$ against the quark-antiquark separation $r$.}
\label{fig:sigma}
\end{figure}

\epsfverbosetrue
\begin{figure}[t]
\begin{center}
\leavevmode
\epsfxsize=2.9in\epsfbox[18 244 592 690]{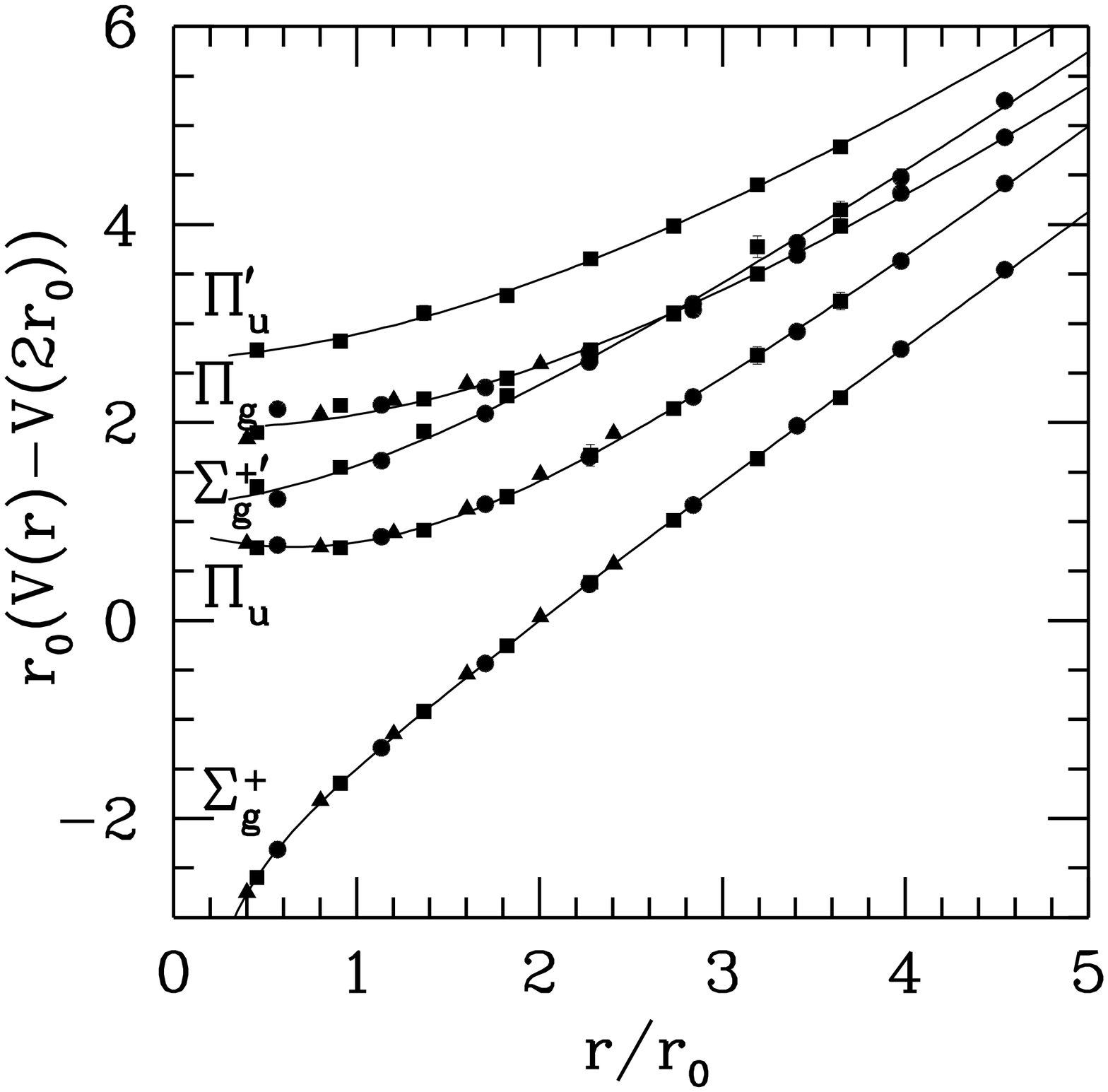}
\end{center}
\caption[figexc]{The static quark potential $V_{\Sigma_g^+}(r)$
 and selected gluonic excitations (see Fig.~\protect\ref{fig:sigma}).
\label{fig:excited}}
\end{figure}

\epsfverbosetrue
\begin{figure}[t]
\begin{center}
\leavevmode
\epsfxsize=2.9in\epsfbox[18 244 592 690]{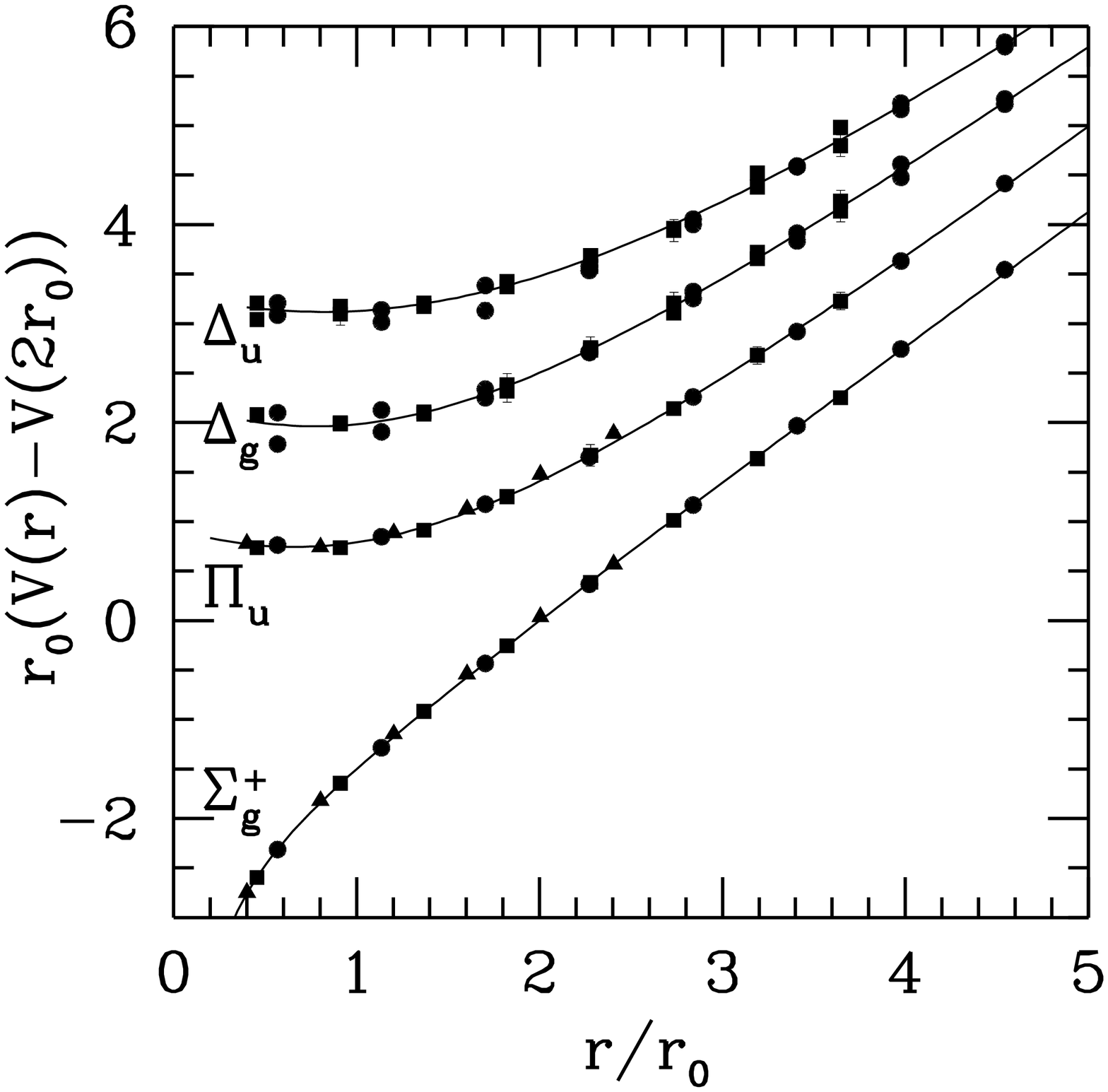}
\end{center}
\caption[figdelta]{The static quark potential $V_{\Sigma_g^+}(r)$
 and selected gluonic excitations (see Fig.~\protect\ref{fig:sigma}).
\label{fig:delta}}
\end{figure}

Monte Carlo estimates of the correlation matrices $W^\Gamma_{ij}(r,\tau)$
were obtained in three simulations.  In each simulation, several
spatial link smearing schemes were used: one scheme was typically
chosen to work well for small $r$, another was optimized for large $r$.
Various run parameters for the three simulations are given in
Table~\ref{table:simparams}; other relevant parameters can be found
in Ref.~\cite{peardon}.  We used the improved action described in
Ref.~\cite{peardon}.  Our use of anisotropic lattices in which the temporal
lattice spacing $a_t$ was much smaller than the spatial spacing $a_s$
was crucial.  Configuration ensembles were generated using a mixture of
Cabibbo-Marinari (CM) pseudo-heat-bath and SU(2) subgroup over-relaxation
(OR) methods. 

The matrices $W^\Gamma_{ij}(r,\tau)$ were reduced
in the data fitting phase to single correlators and $2\times 2$
correlation matrices using the variational method described in
Ref.~\cite{peardon}.  These reduced correlators were fit using a single
exponential and a sum of two exponentials in various ranges $t_{\rm min}$
to $t_{\rm max}$ of the source-sink separation.  The two-exponential
fits were used to check for consistency with the single-exponential fits, and
in cases of favourable statistics, to extract the first-excited state
energy.  Best fit values were obtained using the correlated $\chi^2$ method.
Error estimates were calculated using a $1024$-point bootstrap procedure.

Our results for the static potential and its gluonic excitations
are shown in Figs.~\ref{fig:sigma}-\ref{fig:delta}.  Results from
the $\beta=2.2$, $\beta=2.4$, and $\beta=2.6$ runs are shown using
solid circles, squares, and triangles, respectively.  The results
are expressed in terms of the hadronic scale parameter $r_0$.  The
definition of this parameter and a description of its calculation
are given in Ref.~\cite{peardon}.  The familiar static potential
is shown as $\Sigma_g^+$; the solid curve is a fit to the data using
a Coulomb plus linear form $V_0+e_c/r+\kappa r$.  The curves for all
other potentials are fits using $c_0+\sqrt{b_0 + b_1 r + b_2 r^2}$.
For all $r$ studied, the first-excited potential is the $\Pi_u$;
hence, the lowest lying hybrid mesons must emerge from this potential.

As $r$ becomes very large, the linearly rising $\Sigma_g^+$ potential
suggests that the ground state of the glue may be modelled as a fluctuating
tube or string of colour flux; in this picture, the gluonic excitations
are expected to be phonon-like with energy gaps proportional to $1/r$.
However, it appears that for $r$ below about $1.5$ fm, the gluonic spectrum
cannot be explained in terms of a simple string model.
In Ref.~\cite{hasenfratz}, a QCD motivated bag model was successfully used
to describe both the $\Sigma_g^+$ and $\Pi_u$ potentials for a large range
of $r$.  In this picture, the strong chromoelectric fields of the quark
and antiquark repel the physical vacuum (dual Meissner
effect), creating a bubble inside which perturbation theory is applicable.
In the ground state, the inward pressure on the bubble from the physical
vacuum balances the outward chromostatic force in such a way to produce
a linearly confining potential.  The addition of one or more gluons into the
bag produces the excited potentials; the kinetic energy of the gluons
inside the bubble is a key factor in determining the form of these potentials.
This model has recently been revisited and results (in the ellipsoidal
approximation) for almost all of the potentials studied here are in
remarkable agreement with our findings from the lattice simulations
(see Ref.~\cite{kuti}).

\section{Hybrid quarkonium}
The next step in the BO expansion is to restore
the quark motion by sol\-ving the radial Schr\"odinger equation,
\begin{equation}
\frac{d^2u(r)}{dr^2}+2\mu [E-V_{\rm eff}(r)]\ u(r)=0,
\label{eqn:schrodinger}
\end{equation}
where $V_{\rm eff}(r) = V_{Q\overline{Q}}(r)
 + \langle {\bf L}_{Q\overline{Q}}^2\rangle / (2\mu r^2)$,
$\mu$ is the reduced mass, and $\varphi(r)=u(r)/r$ is the radial
wavefunction.  The total angular momentum of the meson is given by
${\bf J}={\bf L}+{\bf S}$, where ${\bf S}$ is the sum of the spins
of the quark and antiquark, and the orbital factor ${\bf L}=
{\bf L}_{Q\overline{Q}} + {\bf J}_g$, where ${\bf J}_g$ is the total
angular momentum of the glue and ${\bf L}_{Q\overline{Q}}$ is the
orbital angular momentum of the quark and antiquark.  In the 
LBO approximation, the eigenvalues $L(L+1)$ and $S(S+1)$
of ${\bf L}^2$ and ${\bf S}^2$ are good quantum numbers.  The centrifugal
factor is then written as
\begin{equation}
\langle {\bf L}_{Q\overline{Q}}^2\rangle = L(L+1) - 2\Lambda^2
 + \langle {\bf J}_g^2 \rangle.
\end{equation}
For the $\Sigma_g^+$ potential, $\langle {\bf J}_g^2 \rangle=0$.
For the $\Pi_u$ and $\Sigma_u^-$ potentials, we assumed that the one
gluon state was dominant with $\langle {\bf J}_g^2 \rangle=2$.
Mesonic eigenstates of parity and charge-conjugation are linear
combinations of left- and right-handed glue states: $\vert {\rm left}
\rangle + \epsilon\ \vert {\rm right}\rangle$, where $\epsilon=\pm 1$.
Let $\eta=\pm 1$ denote the $PC$ quantum number of the glue.  Then
in the LBO approximation, the parity (P) and
charge conjugation (C) of each meson is given in terms of $L$ and $S$
according to
\begin{eqnarray}
  P &=& \epsilon\ (-1)^{L+\Lambda+1},\\
  C &=& \epsilon\ \eta\ (-1)^{L+\Lambda+S}.
\end{eqnarray}

\epsfverbosetrue
\begin{figure}[t]
\begin{center}
\leavevmode
\epsfxsize=2.95in\epsfbox[18 244 550 680]{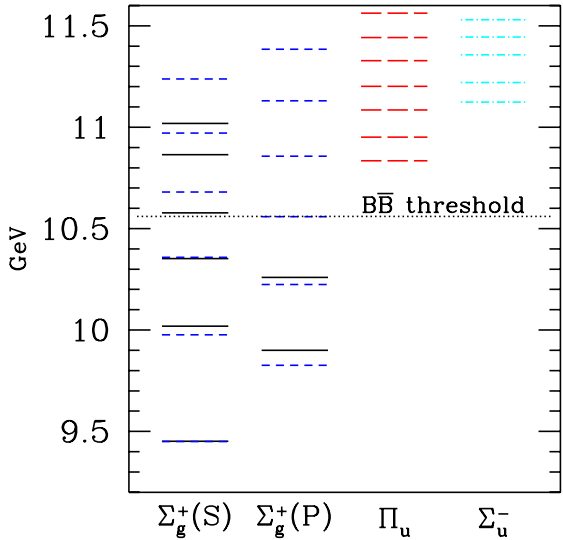}
\end{center}
\caption[figspectrum]{Spin-averaged $b\bar b$ spectrum
 in the leading Born-Oppenheimer and quenched approximations.  Solid lines
 indicate experimental measurements.  Short dashed lines indicate the $S$ and
 $P$ state masses obtained by solving the appropriate Schr\"odinger
 equation in the $\Sigma_g^+$ potential using $r_0^{-1}=0.430$ GeV
 and $M_b=4.60$ GeV for the heavy quark mass. Long dashed and dashed-dotted
 lines indicate the hybrid quarkonium states obtained from the $\Pi_u$
 $(L=1,2)$ and $\Sigma_u^-$ $(L=0,1,2)$ potentials, respectively.
\label{fig:spectrum}}
\end{figure}

\epsfverbosetrue
\begin{figure}[t]
\begin{center}
\leavevmode
\epsfxsize=2.9in\epsfbox[18 244 592 690]{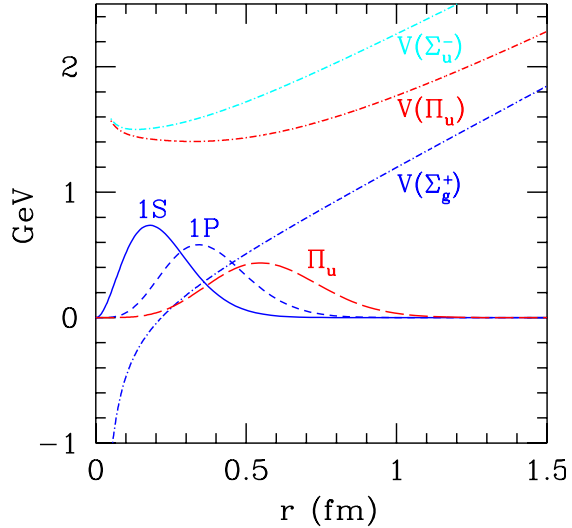}
\end{center}
\caption[figwf]{Static potentials and radial probability densities
 against quark-antiquark separation $r$.  The $\Sigma_g^+$ static
 potential and the $\Pi_u$ and $\Sigma_u^-$ excitations are indicated
 by the dashed-dotted lines.  The solid and short-dashed curves
 indicate the radial probability densities for the $1S$ and $1P$ states,
 respectively, corresponding to the results shown in
 Fig.~\protect\ref{fig:spectrum}. The extended nature of the lowest-lying
 $\Pi_u$ hybrid state is shown by the radial probability density indicated
 by the long-dashed curve.
\label{fig:wf}}
\end{figure}

The potentials computed from Wilson loops in lattice simulations contain
the self-energies of the temporal Wilson lines.  These self-energy
contributions are common to all of the static potentials and must
be removed in order to obtain the $V_{Q\overline{Q}}(r)$ appearing
in Eq.~\ref{eqn:schrodinger}.  Ideally, this can be done by measuring the
$\Sigma_g^+$ potential for very small $r$ and comparing with the running
Coulomb law as predicted from perturbation theory.  In practice, this is
difficult to do.  Instead, we fit our results for the $\Sigma_g^+$ potential
to the form $V_0+e_c/r+\kappa r$; the constant $V_0$ is then our estimate of
the self-energy contributions to be removed.

Results for the $b\bar b$ spectrum are shown in~Fig.~\ref{fig:spectrum}.
The scale was set using $r_0^{-1}\!=\!430$ MeV as suggested from NRQCD
simulations of $b\bar b$ and $c\bar c$ mesons (see Table XX of
Ref.~\cite{peardon}).  The heavy quark mass $M_b$ was tuned in order
to reproduce the experimentally-known $\Upsilon(1S)$ mass: $M_\Upsilon
=2M_b+E_0$, where $E_0$ is the energy of the lowest-lying state in the
$\Sigma_g^+$ potential.  In the LBO approximation, many
mesons are degenerate: the $J^{PC}=0^{-+}$,$1^{--}$ $S$-waves from the 
$\Sigma_g^+$ potential are degenerate; the $0^{++}$,$1^{++}$,$2^{++}$,$1^{+-}$
$P$-waves from the $\Sigma_g^+$ potential have equal masses; states such as
$0^{-+}$,$0^{+-}$,$1^{++}$,$1^{-+}$,$1^{--}$,$1^{+-}$ from the $\Pi_u$
potential are also degenerate.

Below the $B\overline{B}$ threshold, the
LBO results are in very good agreement with the
spin-averaged experimental measurements.  Note that these results make
use of the quenched potentials (which ignore the light quarks) and do not
include spin, retardation, and other relativistic effects.  Above the
threshold, agreement with experiment is lost, suggesting significant
corrections from either the light quarks, relativistic effects, or possibly
mixings between the states from the different adiabatic potentials.  Note
that the mass of the lowest-lying hybrid (from the $\Pi_u$ potential) is
about 10.8~GeV.  Hybrid mesons from all other static potentials are
significantly higher lying.  Above 11 GeV, the LBO approximation based
on the quenched $V_{Q\overline{Q}}$ potentials predicts a very dense
population of hybrid states.  The radial probability densities for the $1S$
and $1P$ conventional states are compared with that of the lowest-lying
$\Pi_u$ hybrid state in Fig.~\ref{fig:wf}.

\epsfverbosetrue
\begin{figure}[t]
\begin{center}
\leavevmode
\epsfxsize=2.9in\epsfbox[18 244 592 690]{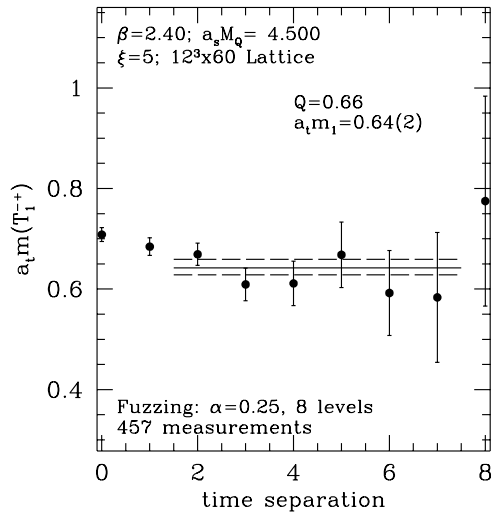}
\end{center}
\caption[fignrqcd]{Effective mass plot showing the results of a
 single-exponential fit to the correlation function of the hybrid
 $T_1^{-+}$ quarkonium state.  The heavy quark propagates according
 to a spin-independent NRQCD action.
\label{fig:nrqcd}}
\end{figure}

\section{NRQCD simulations}
Hybrid quarkonium states may also be studied directly in numerical simulations.
The nonrelativistic formulation of lattice QCD (NRQCD)\cite{nrqcd} is a
particularly efficient means of carrying out such simulations.  We have
recently begun an investigation of the hybrid quarkonium states using
a spin-independent version of the lattice NRQCD action.  The action
included only a covariant temporal derivative and the leading kinetic
energy operator (with two other operators to remove $O(a_t)$ and $O(a_s^2)$
errors); relativistic corrections depending on spin, ${\bf E}$
and ${\bf B}$, and higher derivatives were not included.  The action was
chosen to correspond as closely as possible to the LBO approximation.
In so doing, our treatments of the hybrid mesons using the
Born-Oppenheimer expansion and using NRQCD simulations differed in primarily
two aspects:  lattice spacing errors and retardation effects.  The NRQCD
simulations included retardation effects since the covariant Laplacian 
of the quark kinetic energy operator was treated exactly.  In contrast, the
LBO approximation ignored all retardation effects by
keeping only the leading term in the $1/c$ expansion of the covariant
Laplacian.  A comparison of results from the two approaches should afford
a test of the adiabatic approximation.

Because the signal-to-noise ratio was expected to be poor and the masses of
the hybrid mesons were expected to be large, it was crucial to use an
anisotropic lattice in which the temporal lattice spacing was much smaller
than the spatial spacing.  The operators used to calculate the static
potentials were incorporated into our NRQCD simulation code. We constructed
large sets of gauge-invariant operators in order to minimize excited-state
contamination of our hybrid meson correlators using variational techniques.
We have completed an initial run on a $12^3\times 60$ lattice using an aspect
ratio $\xi=5$ (that is, $a_s=5a_t$).  As the purpose of this run was merely
to discover the quality of signal which was possible, the quark mass was not
tuned and a value $a_sM_b=4.50$ was used.  The effective mass from this run
for the $1^{-+}$ hybrid meson is shown in Fig.~\ref{fig:nrqcd}. 
A convincing plateau was observed from 457 measurements, and we were able
to extract the mass with a $3\%$ statistical uncertainty.  

\section{Conclusion}

A first comprehensive survey of the spectrum of quenched SU(3) gluonic
excitations in the presence of a static $Q\overline{Q}$ pair was presented.
The hybrid quarkonium states were calculated in the leading Born-Oppenheimer
approximation.  The effective mass for the $1^{-+}$ hybrid from a preliminary
NRQCD simulation was also shown.  This work was supported by the U.S.~DOE,
Grant No.\ DE-FG03-90ER40546.

\end{document}